\documentclass[runningheads]{llncs}
\usepackage[T1]{fontenc}
\usepackage{svg}
\usepackage{graphicx}
\usepackage{hyperref}
\usepackage{todonotes}

\newcommand{\reg}[1]{\textsf{#1}}
\usepackage{circledsteps}
\begin{document}
\title{Execution at RISC: Stealth JOP Attacks on RISC-V Applications}
%
%
\author{Loïc Buckwell \and
Olivier Gilles\orcidID{0000-0002-3776-2071} \and \\
Daniel Gracia P\'erez\orcidID{0000-0002-5364-8244} \and \\
Nikolai Kosmatov\orcidID{0000-0003-1557-2813}}
\authorrunning{L. Buckwell et al.}
%
\institute{Thales Research \& Technology, Palaiseau, France \\
 \email{\{loic.buckwell,olivier.gilles,daniel.gracia-perez,\\nikolai.kosmatov\}@thalesgroup.com}
}
\maketitle              
\vspace{-4mm}
\begin{abstract}
  RISC-V is a recently developed open instruction set architecture gaining a lot of attention.
  To achieve a lasting security on these systems and design
  efficient countermeasures, a better understanding of vulnerabilities to novel
  and potential future attacks is mandatory. This paper demonstrates that
  RISC-V is sensible to Jump-Oriented Programming, a class of complex
  code-reuse attacks. We provide an analysis of new dispatcher gadgets we
  discovered, and show how they can be used together in order to build a
  stealth attack, bypassing existing protections. A proof-of-concept attack is
  implemented on an embedded web server compiled for RISC-V, in which we
  introduced a vulnerability, allowing an attacker to remotely read an
  arbitrary file from the host machine.

\keywords{Control-Flow Integrity  \and Code-Reuse Attacks \and Embedded Systems \and RISC-V.}
\end{abstract}
\vspace{-4mm}

\section{Introduction}\label{sec:intro}
The RISC-V Instruction Set Architecture (ISA)\footnote{\url{https://riscv.org}} 
is a novel open Reduced
Instruction Set Computer (RISC) ISA, which is often  
used for embedded systems. While RISC ISAs innately have a
smaller attack surface than Complex Instruction Set Computer (CISC) ISAs, many
of them run critical systems, including industrial control systems 
or
cyber-physical systems,
whose failure may have dramatic consequences
(including environmental disasters and loss of human lives).
Using a novel open ISA brings several benefits. Its novelty brings
security advantages by taking past security failures into experience. Even more
important is the open status, where trust in the architecture relies on
community review. This also enables national independence in microchip supplies,
a very important feature as target systems may be strategical, and export
restrictions become more common.

While most RISC-V architectures offer a satisfying level of security compared to
similar classes of systems, they will increasingly become the target to complex
attacks as their relevance in the industrial and strategical field increases.
Eventually, state-backed attackers are deemed to attack them.
In order to anticipate this threat, security researchers face the challenge to
anticipate potential vulnerabilities and imagine suitable protection mechanisms.
Code-Reuse Attacks (CRA), and specifically those based on Jump-Oriented Programming (JOP), are
among the most complex attacks to realize, but also to prevent. They can be very
powerful when successful, as they can allow the attacker to run an arbitrary
sequence of instructions within the corrupted application. In this article we
adopt the attacker's point of view and try to perform a JOP attack, with the
intent of (1) better understanding the vulnerabilities of RISC-V systems, and
(2) ultimately designing better countermeasures to prevent these attacks.

\emph{Contributions.} We summarize our contributions as follows:
\begin{itemize}
\item a first analysis of vulnerabilities to JOP attacks on RISC-V architecture;
\item a description of new dispatcher gadgets enabling JOP attacks to bypass modern mitigations on RISC-V architecture;
\item a demonstration of feasibility by implementing and testing a stealth JOP attack on a vulnerable RISC-V application.
\end{itemize}

\emph{Outline.}
Section~\ref{sec:background} introduces code-reuse attacks, countermeasures
against them and the limitations of the latter. Section~\ref{sec:riscv-jop}
briefly describes the \mbox{RISC-V} ISA attack surface regarding JOP attacks, and
introduces a new kind of gadgets, increasing gadget availability above the level
of ROP attacks. Section~\ref{sec:experiment} describes an attack we developed
against a vulnerable RISC-V application using techniques described in previous
sections, and how we use them to reach a new property for CRA, stealth.
Section~\ref{sec:related} compares our approach to other efforts related to
Jump-Oriented Programming and RISC-V security. Finally,
Section~\ref{sec:conclusion} provides a conclusion.

\section{Code-Reuse Attacks Overview}\label{sec:background}

The aim of a \emph{Code-Reuse Attack} (CRA) is to take control of an application
execution through the use of existing code within the target application in
order to perform unintended or malicious actions.

A CRA is not in itself a primary attack, but relies on an earlier memory
corruption allowing to hijack the execution flow. Such vulnerabilities are
well-known, but still prevalent in many systems~\cite{Younan04codeinjection}.
The originality of CRAs comparatively to regular code injection is that, instead
of redirecting the execution flow toward injected code (generally, a shellcode
in a buffer), it redirects the execution toward existing code in the
application in order to obtain a malicious effect. A simple example of such
attacks is return-to-libc~\cite{Solardesigner97}, where the execution flow is
redirected to a single function after manipulation of arguments within the
stack of the corrupted function. More sophisticated attacks with the same
principle of stack corruption have emerged, among which the Return-Oriented
Programming (ROP) technique~\cite{Shacham07,Carlini14}. It consists in chaining
\emph{gadgets}, i.e. code snippets composed of a few instructions and ending with a
linking instruction. In the case of ROP, the linking instruction is a return to a
caller instruction which pops the next gadget address from the corrupted stack,
handing on execution flow to it, and so on. Using this approach, the attacker
can run an arbitrary sequence of legit instructions, effectively running a
malicious action using the target application code.

\subsection{Countermeasures}\label{sec:background:countermeasures}

Multiple methods were proposed and used in order to defend against
return-to-libc and ROP.
Address Space Layout Randomization (ASLR) randomize base addresses of memory
mappings. The attacker needs to guess the address of target function or gadgets.
Stackguard~\cite{Cowan98} introduced the notion of canaries to protect the
integrity of the stack. Yet solutions relying on secrets depend much on the
system entropy, which tends to decline as the system uptime increases --- an
important issue for embedded systems that can run for decades without reboot.
Both ASLR and Stackguard are even weaker on 32-bit
systems~\cite{Shacham05}, and several techniques have been proposed to bypass
them.

Abadi et al.~\cite{abadi05} first formally identified a process property named
\emph{Control-Flow Integrity} (CFI), defined by the adherence of the runtime execution
flow to its intended behavior. In order to ensure this property against
attackers, they proposed two complementary protections: shadow stack and landing
pads\footnote{A specification of shadow stack and landing pads for RISC-V is
currently under ratification~\url{https://github.com/riscv/riscv-cfi/}}.
\emph{Shadow stack} protects backward-edge jumps by pushing procedure return addresses
to a memory protected stack at call time. When a procedure returns,  its return
address is popped from both stacks and compared. If they differ, a memory
corruption is detected.
\emph{Landing pads} are special instructions protecting forward-edge jumps. When
implemented, each jump destination must be one of these instructions. However,
they rely on the compiler generating them, and even if an application is
compiled with them, it can still use shared libraries that do not
use landing pads, effectively loosing benefits for corresponding code.
Nevertheless, these protections make theoretically all kinds of CRA nearly
impossible to implement, and do indeed stop most
return-to-libc and ROP attacks, although often leading to significant fall of 
performances~\cite{Burow17}.

\subsection{Overcoming the Shadow Stack: Jump-Oriented Programming}\label{sec:background:overcoming-shadow-stack}

\begin{figure}[tb]
  \begin{center}
    \includegraphics[scale=0.55]{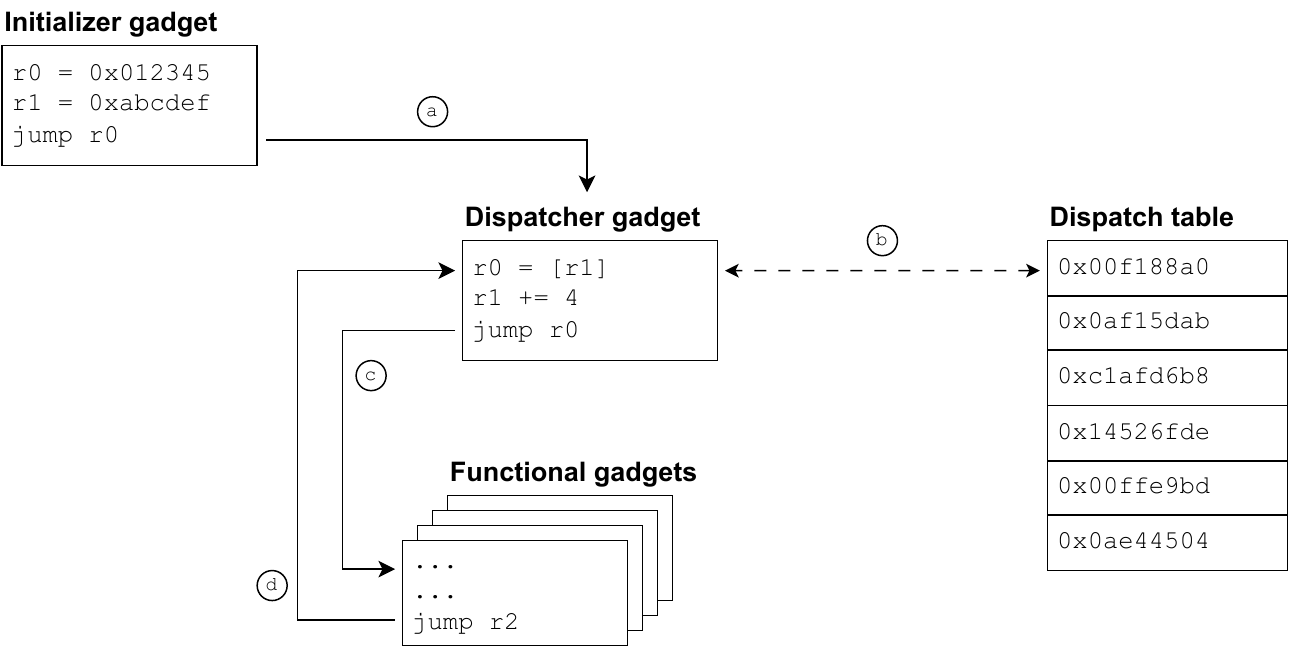}
  \end{center}
\vspace{-4mm}
  \caption{JOP mechanic principle.}\label{fig:jop}
\vspace{-4mm}
\end{figure}

Much like ROP, JOP consists in assembling \emph{functional gadgets} containing
useful instructions present in the target application in order to perform 
a malicious action. 
However in the case of JOP, the chaining
mechanism must be done by a \emph{dispatcher gadget}. Its role is to load and
jump to the next {functional gadget} from a \emph{dispatch table},
generally injected into a buffer. Each {functional gadget} must then end
with a jump to the {dispatcher gadget}. To do this, at least two registers
need to be reserved: one for the dispatcher gadget and one for the dispatch
table. The \emph{initializer gadget} is responsible to set them to their
respective addresses. Figure~\ref{fig:jop} illustrates this mechanism
with an  example, where  \reg{r1} and \reg{r2} are
reserved registers (respectively, for the dispatch
table and the dispatcher gadget), and \reg{r0} is used to branch to functional gadgets.

Assembling a JOP gadget chain is far more complex than doing so for a ROP gadget
chain for multiple reasons.
First, reserved registers must be set to appropriate values prior to passing
execution flow to the dispatcher gadget. This must be done with another
specialized gadget, which is only executed once, and is referred to as the
\emph{initializer gadget}. The dispatcher and initializer gadgets are thus intimately
tied, as they must work with the same reserved registers. Although the
initializer and dispatcher gadget patterns are quite simple, there are few of
them in practice, and viable combinations of both are even more scarce. In
addition to this difficulty, for any viable pair, there must be enough
compatible functional gadgets, i.e. ending with a jump to the dispatcher gadget
register in order to build the actual attack code. For this reason, the best
dispatcher gadget register is the one for which we find the most functional
gadgets. In practice, most common functional gadgets are procedure epilogues but
using them would trigger shadow stack detection if implemented. Other registers,
e.g. argument registers, are hard to use as reserved registers. Reserving them
would make argument passing complex, reducing the attack effectiveness.
Side-effects in functional gadget must also be considered as they break the
gadget chain management by clobbering the reserved registers.
For these reasons, JOP remains mostly theoretical. The most prominent example
publicly known to this day is a complex rootkit targeting
x86/Linux2.6~\cite{Cheng11}.

Since searching and chaining gadgets is highly dependent on both ISA and 
Application Binary Interface (ABI),
its feasibility in new architectures is not proven. In the following sections, we
demonstrate the feasibility of JOP attacks on applications compiled for the
RISC-V architecture, and introduce a new kind of dispatcher gadget enabling the
use of procedure epilogues as functional gadgets without triggering shadow stack
detection, allowing to craft stealth attacks with a greater attack surface.

\section{Bringing JOP to RISC-V}\label{sec:riscv-jop}

\subsection{Attack Surface}

Ability to perform JOP attacks for a given architecture heavily relies on
gadget availability. As such, they represent the attack surface. We used
RaccoonV\footnote{\url{https://github.com/lfalkau/raccoonv}}, an open-source tool 
to search for RISC-V JOP gadgets, to gather statistics about available gadgets 
for some applications. Table~\ref{tab:rvstat} shows the number of available 
gadgets per register in the GNU libc 2.34 compiled for RISC-V 32 bits with M, A 
and C extensions (RV32IMAC).

\begin{table}[tb]
	\begin{center}
	\begin{tabular}{|l|l|l|l|l|l|l|l|l|l|l|l|l|l|l|l|l|l|l|l|l|l|l|l|l|l|l|l|l|l|l|l|}
		\hline
                Register & \reg{ra} & \reg{e5} & \reg{t1} & \reg{t3} & \reg{tp} & \reg{a4} & \reg{s0} & \reg{s2} & \reg{a2} & \reg{a0} & \reg{sp} & \reg{s1} & \reg{a3} & \reg{t5} & \reg{s8} \\ \hline
		Available gadgets & 4557 & 810 & 318 & 255 & 239 & 184 & 183 & 157 & 147 & 106 & 97 & 86 & 83 & 79 & 68 \\ \hline
	\end{tabular}
	\caption{Gadget availability per register in libc (top 15).}\label{tab:rvstat}
    \vspace{-4mm}
	\end{center}
\end{table}

\subsection{Introducing Autonomous Dispatcher Gadgets}

Table~\ref{tab:rvstat} shows that \reg{ra} and \reg{a5} are the registers of
choice for the dispatcher gadget register using a classic JOP scheme. However,
jumping to \reg{ra} from functional gadgets would trigger shadow stack
detection if implemented. On the other hand, \reg{a5} tends to be used by GCC
when calling function pointers. In our experience, as most dispatcher gadgets
are generated by the call of function pointers in a loop, they already use
\reg{a5} to jump to functional gadgets, making them impractical candidates.
In practice we did not find any suitable combination of reserved
registers allowing to chain system calls in a convenient
way~\cite{gilles2022}.

\begin{figure}[tb]
	\begin{center}
		\includegraphics[scale=0.5]{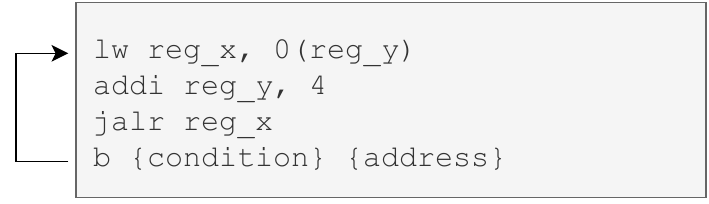}
  \vspace{-4mm}
		\caption{Autonomous dispatcher gadget (ADG) example.}\label{fig:dispatcher_gadget}
  \vspace{-4mm}
	\end{center}
\end{figure}

While searching for more suitable gadgets in RISC-V applications, we found a
new kind of a dispatcher gadget, illustrated in
Figure~\ref{fig:dispatcher_gadget}, that we call an \emph{autonomous dispatcher gadget}
(ADG).

It has two differences from a classical dispatcher gadget. First, the
instruction used to link to functional gadgets is a JALR, instead of a JR. In
order to use such dispatcher gadget without triggering shadow stack detection,
functional gadgets ending with a jump to \reg{ra} now can --- and must ---  be used.
Doing so, each round-trip between the dispatcher gadget and functional gadgets
will correspond to a legit procedure call.
However, when using a JALR instruction to link to functional gadgets, \reg{ra}
will be set to the next instruction in the dispatcher gadget, instead of the
dispatcher gadget address itself. While this would be an issue with a classical
dispatcher gadget, the second difference of the ADG is that it links back to
itself when getting control back, which justifies its name. However, in each ADG we found
or managed to reproduce, this self-linking instruction was conditional. While
the attacker has to ensure the condition remains true in order to preserve the
chain, it is in practice easily feasible as discussed in Section~\ref{sec:experiment}.

\subsubsection{Code pattern}

\begin{figure}[tb]
  \begin{center}
    \includegraphics[scale=0.5]{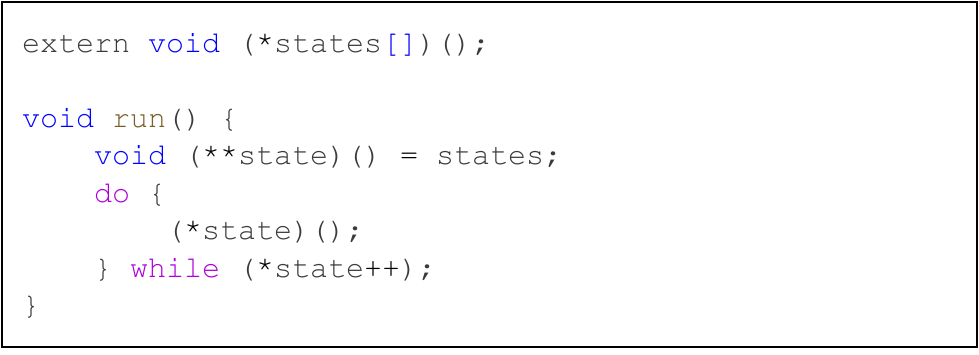}
  \end{center}
  \vspace{-4mm}
  \caption{Dispatcher gadget code pattern.}\label{fig:adg_code}
  \vspace{-4mm}
\end{figure}

The first autonomous dispatcher we found was in the GNU libc 2.34, compiled with
the second GCC level of optimization (\textit{-O2}). We also managed to generate
several autonomous dispatcher gadgets with simple and realistic code patterns,
each involving function pointer calls inside a loop. Figure~\ref{fig:adg_code}
is one of them.

\subsubsection{Reserved registers}

In order to craft a JOP attack using an ADG, only one reserved register is
required: the dispatch table register. Indeed, the ADG links back to itself with
a JAL instead of a JALR, which would  require another reserved register as
classic dispatcher gadgets.

As a consequence, the initializer can use any register to jump to the dispatcher
gadget the first time.
The only register it must share with the ADG is the dispatch table register.
Reducing the reserved registers constraint between these two gadgets considerably increases
 the probability to find a compatible pair.

From our experience, generated ADGs often use first available saved registers
(\reg{s0--11}) as the dispatch table register, which is convenient because there
is a good balance of gadgets loading them from the stack (potential initializer
gadgets) and gadgets which do not clobber them, making them compatible
functional gadgets.

\section{Attacking Real-World RISC-V Applications}\label{sec:experiment}

In order to show the feasibility of JOP attacks in the RISC-V
architecture, we decided to test it on an embedded application: Mongoose
web server, and more precisely the provided http-server, in which we
introduced a memory corruption vulnerability.

Mongoose web server is a target of choice because it can be interacted with
from the network. While its features would probably be restricted in real-life
scenarios, it is still very relevant to demonstrate JOP feasibility. The aim of the
attacker in our scenario is to remotely read the private part of the target's
root SSH key, which is base64-encoded and stored on disk.

Mongoose is developed in C. We compiled it for RISC-V 32 bits using GCC,
disabling Position-Independent Executable (PIE). It aims to run over a Linux
system, and dynamically links to the libc. In order to keep the attack as
portable as possible, we only relied on the application itself for the memory
corruption and execution-flow hijacking. Every gadget used then has been found
on the libc, which has been compiled with modern binary protections, using
\emph{-fstack-protector-all -D\_FORTIFY\_SOURCE=2}, and the second level of
optimization (\emph{-O2}). Both operating system and target application were
executed and validated on a RISC-V CVA6 32-bit softcore
design\footnote{\url{https://github.com/openhwgroup/cva6}}~\cite{zaruba2019cost}
running Linux, deployed on a Genesys2 FPGA.

In the following subsections, we explain how we were able to steal the key
stored on disk using exclusively JOP thanks to an ADG, by crafting a malicious
HTTP request.

\subsubsection{Attack Model}

\begin{figure}[tb]
	\begin{center}
		\includegraphics[scale=0.38]{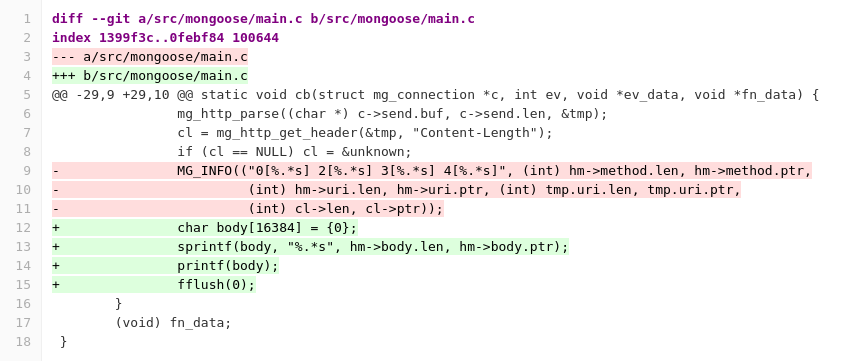}
  \vspace{-4mm}
		\caption{Mongoose introduced vulnerability diff.}\label{fig:vuln-diff}
  \vspace{-4mm}
	\end{center}
\end{figure}

The attack we realized in our experiment relies on a memory vulnerability, for
instance a format string vulnerability, allowing to highjack the execution flow
of the target application. We did not activate ASLR on the target application,
since it can be bypassed by different techniques, and is out of scope of this
research~\cite{Shacham05}. Figure~\ref{fig:vuln-diff} shows the diff of the
original Mongoose code with the vulnerability we introduced.

We also made the hypothesis that, during the reconnaissance phase of the attack,
the attacker is able to access an exact twin of the target application (either
by rebuilding it with the same options and environment, or by acquiring an
device running said application), and run it with a debugger. Thus, potential
anti-debug, anti-reverse or anti-tampering measures are not considered within
the scope of our experiment.

\subsubsection{Attack overview}

The crafting of the attack consists in five steps divided between the two first
stages of the cyber kill chain: reconnaissance and weaponization.

The three steps involved in the reconnaissance stage are:
\begin{itemize}
	\item identification of a memory vulnerability;
	\item identification of available gadgets in the application;
	\item definition of the attack aim.
\end{itemize}
The two steps involved in the weaponization stage are:
\begin{itemize}
	\item crafting of the JOP chain;
	\item crafting of the malicious payload.
\end{itemize}

As a first step the attacker must identify a memory vulnerability allowing to
hijack the execution flow toward the initializer gadget. This step is not covered
in detail in this document, but many techniques and tools exist in order to
identify such vulnerabilities~\cite{Younan04codeinjection}.

In our experiment, we inserted a format string vulnerability within the target
application, allowing an attacker to perform arbitrary writes. We used it to
overwrite the Global Offsets Table (GOT) entry of the fflush function of the 
stdio C library --- which is called right after the vulnerable printf.

\subsubsection{Identifying gadgets}

\begin{figure}[tb]
  \begin{center}
    \includegraphics[scale=0.45]{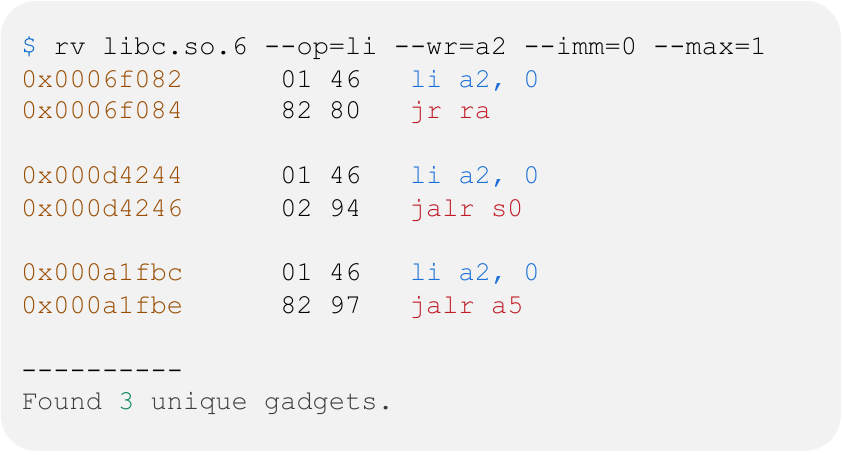}
  \end{center}
  \vspace{-4mm}
  \caption{RaccoonV output with a query on libc.}\label{fig:rv_sample}
  \vspace{-4mm}
\end{figure}

The identification of available gadgets in the application is the most important
stage in the attack, as it decides which assets can be targeted by the attack.
Too few, or not diverse enough gadgets will restrict the range of available
targets in the best case, or make the attack impossible in the worst case. In
our experiment, we searched for gadgets using RaccoonV, an open-source tool we
built in order to automate the gadget discovery based on queries.
Figure~\ref{fig:rv_sample} shows an output example of a simple query using RaccoonV.

In this example, we queried for gadgets in the libc, that loads the immediate 0
(\texttt{{-}-op=li {-}-imm=0}) in the register \reg{a2} (\texttt{{-}-rr=a2}),
and is at most 1 (\texttt{{-}-max=1}) instruction long (excluding the gadget 
linker i.e. the final JALR instruction).
The attacker can then focus on the gadget offsets (left column of the output).

\begin{figure}[tb]
  \begin{center}
    \includegraphics[scale=0.5]{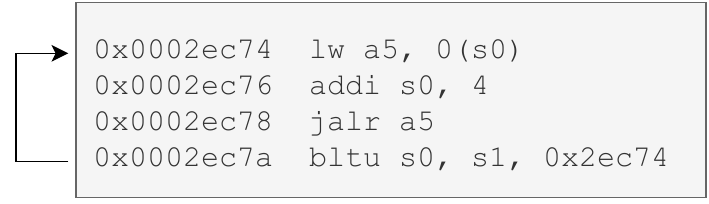}
  \end{center}
  \vspace{-4mm}
  \caption{Autonomous dispatcher gadget found in libc.}\label{fig:dispatcher_gadget_libc}
\end{figure}

As the dispatcher gadget is among the hardest to find, it is strongly
encouraged to find one first, and to build the attack around. After a first
analysis, we found 10 potentially viable dispatcher gadgets in the libc, and
analyzed each by hand. Luckily enough, we found an ADG, described in
Section~\ref{sec:riscv-jop} and illustrated in
Figure~\ref{fig:dispatcher_gadget_libc}. Its self-linking instruction is
conditional, and in order to use it without breaking the chain, we must ensure
\reg{s0} remains inferior than \reg{s1}.

\begin{figure}[tb]
  \begin{center}
    \includegraphics[scale=0.5]{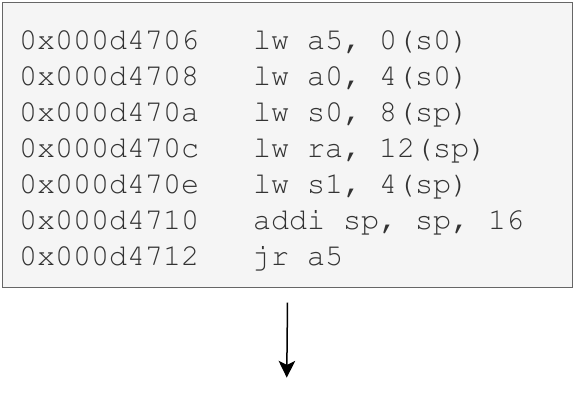}
  \end{center}
  \vspace{-8mm}
  \caption{Initializer gadget found in libc.}\label{fig:init_gadget_libc}
  \vspace{-4mm}
\end{figure}

Finding an initializer gadget in the libc which allows setting \reg{s0} and
\reg{s1}, and does not use \reg{ra} as the linking register was a difficult
task, but thanks to the ADG, more candidate were
available and we managed to find one, illustrated in
Figure~\ref{fig:init_gadget_libc}. It has two side-effects: the loading of \reg{ra}
from the stack and the stack pointer increment. While the former has no
importance, the latter needs to be fixed later with functional gadgets in order
for the attack to remain stealth. Values loaded from the stack pointer
\reg{sp} and the frame pointer \reg{s0} can be arbitrarily set through the
format string vulnerability before handling the execution to the initializer
gadget.

Using these two gadgets to build the JOP attack allows using functional gadgets
ending in \reg{ra}, which are the most common. For instance, out of 7915 unique
gadgets found in the libc with RaccoonV, 4557 end in a jump to \reg{ra}.

\subsubsection{Definition of attack objective}

The objective of the attack is to be decided from (1) system assets accessible
by the application and (2) available functional gadgets identified. The first
part requires acquiring a deep understanding of the target application and system.
For the second part, binary access and static analysis is sufficient.

From our experience, having a large number of compatible functional gadgets (4557 in our
case) gives the attacker enough freedom to build advanced attack codes as long as
the attack does not involve complex code patterns such as loops or conditions.

\begin{figure}[tb]
  \begin{center}
    \includegraphics[scale=0.5]{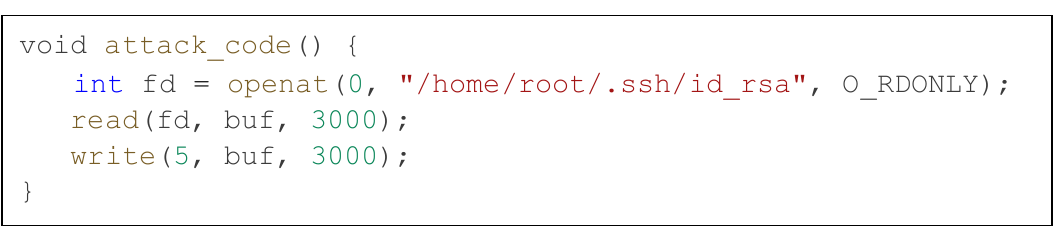}
  \end{center}
  \vspace{-6mm}
  \caption{C formalization of the attack objective.}\label{fig:attack_code}
  \vspace{-2mm}
\end{figure}

In our experiment, our objective is to read the root private SSH key file used
to administrate the server without being detected. Knowing this key allows a
third party to perform any action at the most privileged level on the remote
machine, such as stealing personal or confidential information, performing
website defacing, but also installing more advanced, and potentially persistent malware. We define the attack code as the
C code equivalent shown in Figure~\ref{fig:attack_code}.

The first \emph{openat} argument is ignored when the path is absolute. We used
5 as the file descriptor to write the key as it turned to be the first file
descriptor assigned to clients by the HTTP server. If it is free when the
attacker sends the malicious request, the attacker's request will be assigned it.

\subsubsection{Design of the JOP chain}

\begin{figure}[tb]
  \begin{center}
    \includegraphics[scale=0.4]{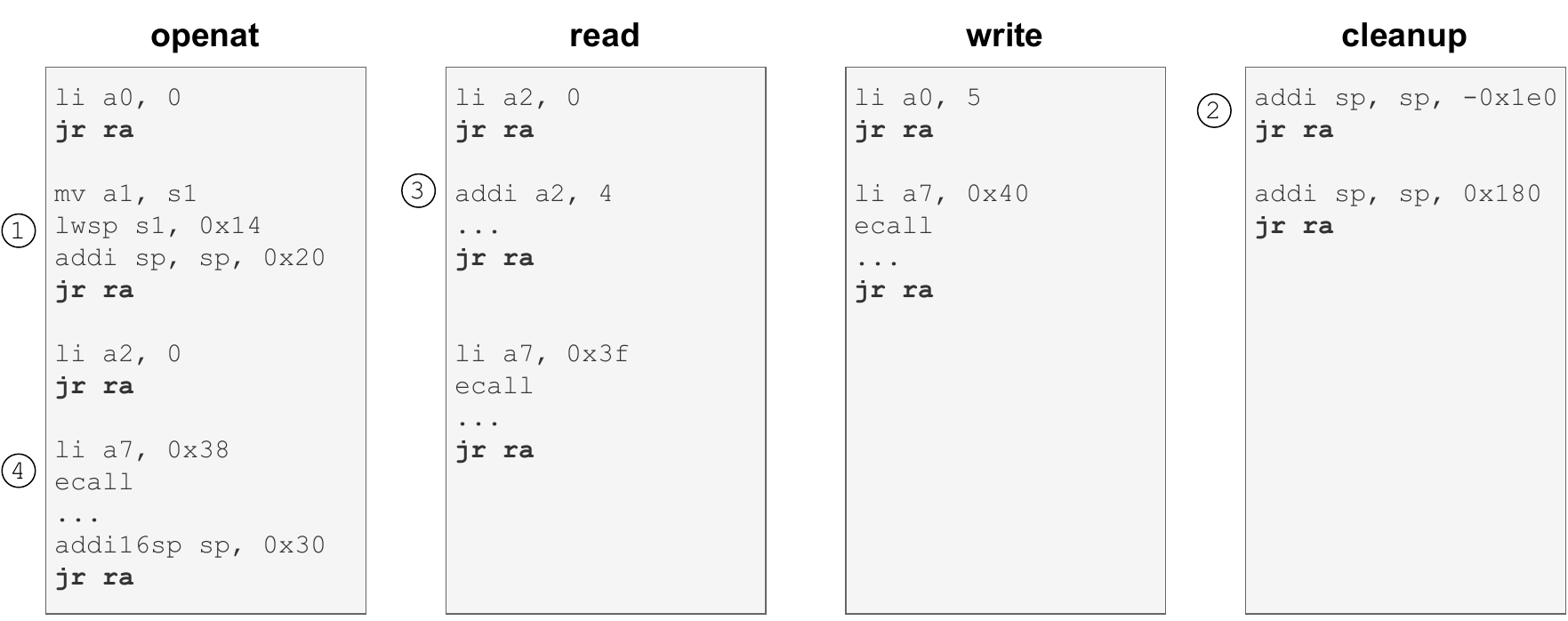}
  \end{center}
  \vspace{-6mm}
  \caption{JOP chain.}\label{fig:jopchain}
  \vspace{-4mm}
\end{figure}

Once the functional gadgets are identified and the objective of the attack is
defined, the actual gadget chain can be crafted. In our case, it consists in a
chain of 3 syscalls and a cleanup step, in order to be able to return to
original code to the original execution flow without being detected (e.g. by 
making the program crash or being detected by the shadow stack). Using RaccoonV, 
we managed to build the distgadget chain shown in Figure~\ref{fig:jopchain}, in which ``\ldots'' mark instructions which are not useful in the attack (but still have side effects).

Some gadgets have side-effects e.g. modifying' the stack pointer or loading
registers from the stack \Circled{1}. While the former is fixed by the cleanup
part of the JOP chain, the latter will need to be considered while forging our
payload. The gadget that subtracts \reg{sp} \Circled{2} is not present in the
original code: it is a valid instruction starting at an unexpected offset,
often called shifted offset of misaligned instruction. For read and write
syscalls, we need \reg{a2} to be big enough to process the whole file
($\approx$3kb). The most appropriate gadget we found is one that increments
\reg{a2} by 4 \Circled{3}, we used this gadget 651 times to achieve this goal.
For each system call, we found gadgets in the libc that set \reg{a7} to the
right identifier, and performs the syscall e.g. the \emph{openat} one \Circled{4}.

At the end of the dispatch table, we also added the address of the original code
we want to jump back to. This address will be called by the dispatcher gadget as
any functional gadget, so we must return to the code at some point where it will 
not return beyond this new stack entry to avoid being detected by the shadow 
stack. In our case, we decided to jump back to the http-server main loop. 
However, to return to this point without making the program crash, we must 
reset some registers at their expected values. To do so, we used the epilogue 
of the function we hijacked as a last gadget before returning to original code.

At the end of the dispatch table, we also added the address of the original code
we want to jump back to. This address will be called by the dispatcher gadget as
any functional gadget so we must return to the code at some point where it will not
return beyond the resulting shadow stack entry. In our case, we decided to jump
back in the http-server main loop. However, to return to this point without
making the program crash, we must reset some registers at their expected values.
To do so, we used the epilogue of the function we hijacked as a last gadget
before returning to original code.

\subsubsection{Initialization of the attack}

\begin{figure}[tb]
  \begin{center}
    \includegraphics[scale=0.45]{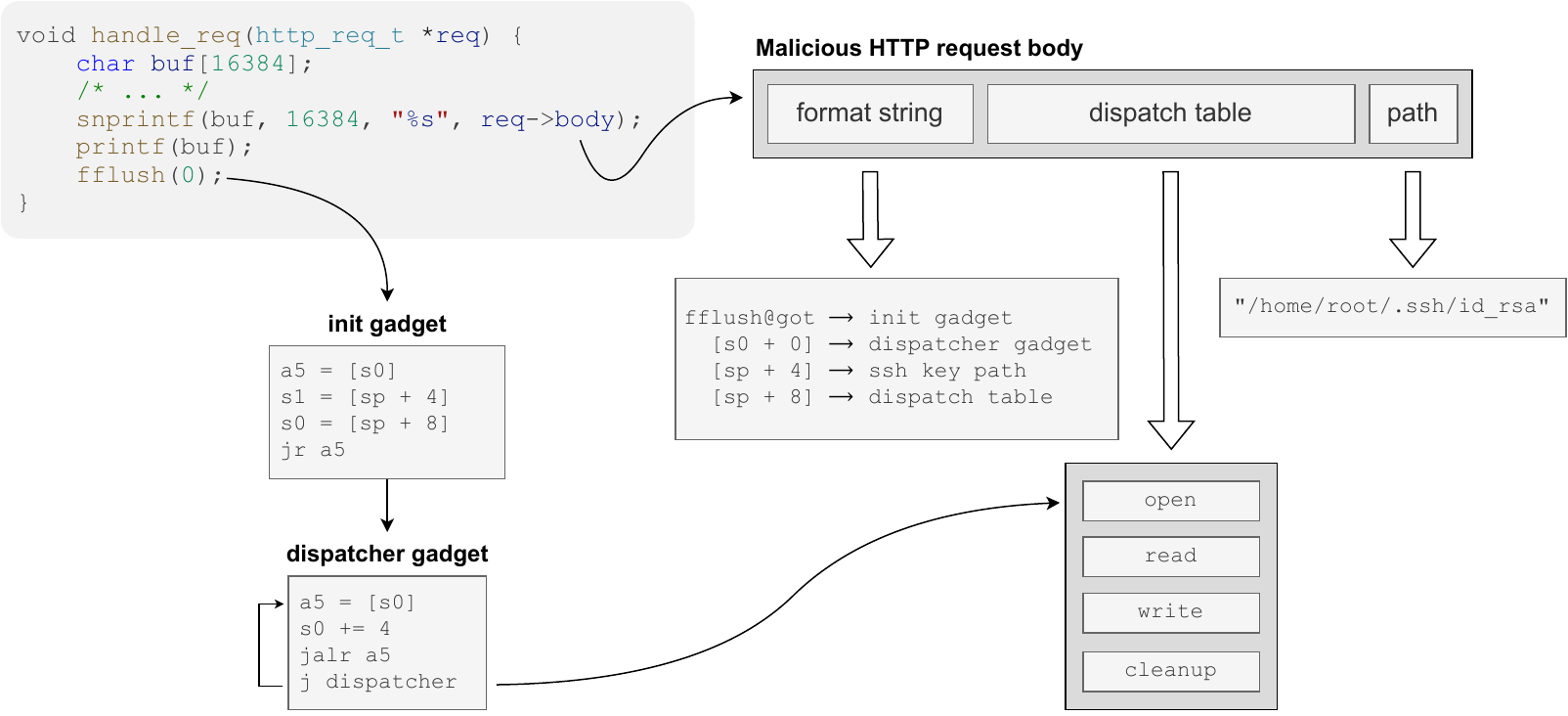}
  \end{center}
  \vspace{-4mm} 
  \caption{Mongoose stealth JOP attack.}\label{fig:mongoose_attack}
  \vspace{-4mm} 
\end{figure}

Once the chain has been designed, it can be encoded in the dispatch table. We
use a sequential table, which addresses all the gadgets, including repetitions
of the same gadgets like in the case of incrementing \reg{a2}, and ends with
the address we want to return to in the original code. We can then forge our
malicious HTTP POST request body that will trigger the format string
vulnerability to set addresses that will be assigned to registers by the
initializer gadget, and overwrite the fflush GOT entry, effectively redirecting
execution-flow toward the initializer gadget. We also append the dispatch
table, and the path of the file we want to read next to the format string. In
the end, the request body is 3075 bytes long, which is acceptable for a POST
request. Figure~\ref{fig:mongoose_attack} illustrates the fully-fledged JOP
attack.

While not related with JOP, in order for the attack to be stealth, we need to
restore the fflush GOT entry we modified. To do so, a second HTTP request
triggering the same format string vulnerability does the job. We can either
patch it with the actual fflush address if we know it, or set it back to the
default Procedure Linkage Table (PLT) stub address, to let the dynamic linker resolve 
its address again at the next invocation.

\subsection{Results and Limitations}

By using the techniques presented in this section, we were able to steal the
private SSH key of the root user stored on target's disk. While not implemented
in our target processor the attack should not trigger the shadow stack
detection. The server still runs fine after the attack, and other clients can
still interact with it normally.

As of today, it seems that landing pads could not be defeated with these
techniques. To do so, we would need to use bigger gadgets --- and eventually
full functions --- which may become impractical.

\subsection{Next Steps}


As far as we know, there is no publicly available implementation of a standalone
RISC-V shadow stack, without other CFI mitigations such as landing pads. For
this reason, while we can theoretically bypass it, we were not able to test our
attack against an actual shadow stack implementation. This is left as future work.

\section{Related Work}\label{sec:related}

\subsection{Building JOP Attacks}

Brizendine et al.~\cite{joprocket21} proposed and implemented a method allowing
building JOP gadget chains for the x86 architecture. The method
relies on predefined gagdets of known characteristics, found in Microsoft
Foundation Class (MFC). While this approach can reliably build JOP chains when
known libraries are involved, it implies to update the tool catalog whenever
these libraries are updated, and to perform in-depth analysis of them. 

Other
approaches try to build partial gadget chains by analyzing the whole used
code, binary and libraries~\cite{Vishnyakov21,Nurmukhametov21}. While
some can integrate sub-chains of JOP gagdets within a ROP chain, none of them
can build full JOP chains to our knowledge, making them easily detected by
ROP-targeted countermeasures such as the shadow stack. Other tools able to help
building JOP chains on RISC-V include ROPgadget\footnote{\url{https://github.com/JonathanSalwan/ROPgadget}}
and radare2\footnote{\url{https://github.com/radareorg/radare2}}, which can search JOP
gadgets but not build gadget chains on RISC-V architectures, as they have no method
for discovering dispatcher gadget or initializer gadget. They are primarily designed
to build ROP chains.

Gu et al.~\cite{gu2020riscvrop} identified a specific pattern of instructions
allowing linking functional gadgets in RISC-V architectures, introducing the
concept of ``self-modifying gadget chain'' to save and restore register values in
memory. They also demonstrated the Turing-completeness of their solution. Adapting
self-modifying gadget chain to JOP is indeed a promising solution to increase our
capacity to build effective gadget chains. Jaloyan et al.~\cite{Jaloyan20} reached
the same result by abusing compressed instructions (\emph{overlapping}). Our attack
also uses this approach, and applies it to JOP attacks.

Trampolines-based approaches are somewhat a missing link between ROP and JOP.
A trampoline itself (an update-load-branch suite of instructions) is the 
ancestor of the dispatcher gadget and, instead of exploiting an arbitrary 
memory, uses hardware-maintained registers such as \reg{ra} (return address 
register) to jump to the next functional gadget~\cite{Checkoway10}. While they 
do not rely on return-specific instructions (which do not exist in RISC-V 
anyway), they do imply that large segments of the stack need to be corrupted, 
hence making them vulnerable to stack canaries and the shadow stack. 
Erd\"odi~\cite{Erdodi13} proposed 
a solution to find classical dispatcher gadgets on x86 for different operating
systems. As they are scarce, and trampolines patterns tend to be more common,
the latter are still used~\cite{Sadeghi14}. In addition to providing a solution
in RISC-V architecture for JOP gadget chaining, our discovery of the ADG
greatly increases the number of available JOP gadgets, effectively making them 
as common as ROP gadgets and eliminating the need for trampolines.

\subsection{Defenses from CRA}

Austin et al.~\cite{morpheus21} published the MORPHEUS II solution for RISC-V. This
hardware-based solution aims at defeating memory probes trying to bypass address
randomization by providing a reactive, fine-grain, continuous randomization of
virtual addresses, as well as encryption of pointers and caches. This solution,
while having a low overhead in terms of energy consumption and area, is quite
intrusive in the hardware and may require efforts for certification in critical
applications. While authors make no claim about stopping JOP attacks, probe-resistant
ASLR may be difficult to bypass for an attacker.

Palmiero et al.~\cite{Palmiero18} proposed a hardware-based adaptation of Dynamic
Flow Information Tracking (DIFT) for RISC-V, with the ability to detect most function
pointers overwriting, whether directly or indirectly, and in any memory segment, thus
allowing blocking the attack at its initialization stage. Although this approach seems
indeed powerful, it implies modification of RISC-V instructions behavior in I and M 
extensions for RISC-V 32-bits, as well as in the memory layout (by adding a bit every 
8 bits of memory). Such modifications, in addition to drifting away from the RISC-V 
ABI, are likely to make certification difficult, a serious drawback in critical 
industrial systems. De et al.~\cite{heapsafe22} implemented a chip compliant to RISC-V, including a Rocket Custom Coprocessor (RoCC) which extends the RISC-V ISA with new 
instructions allowing safe operation on the heap. The authors ensure heap size 
integrity and prevent use-after-free attacks, at the cost of an increase of 50\% 
of average execution time on their benchmarks.

\section{Conclusion}\label{sec:conclusion}

Anticipating security vulnerabilities for RISC-V systems in order to identify
and prevent possible attacks is an important challenge.
Building attacks is a necessary step to test platforms and detect application
vulnerabilities, as adversary actors (\textit{black hat} hackers) will
eventually find them out.
In this article, we contribute by demonstrating the feasibility and
a practical way to realize {jump-oriented programming} (JOP) attacks, allowing for
more extensive security testing.

We have introduced a new variant of dispatcher gadget, the \emph{autonomous
dispatcher gadget} (ADG), which greatly improves the RISC-V JOP attack surface
by enabling the use of ROP gadgets.
While its rigorous validation against a CVA6 implementing a shadow stack is left as future work,
we are convinced that it will be able to bypass shadow stack mitigation.

We have demonstrated a JOP attack on a RISC-V platform using a real world
application, the Mongoose web server,  commonly used in embedded
critical systems.
After adding a single memory vulnerability, we 
were able to take control of the application in order to perform an adversary
action, sending a private key to a remote attacker.
Thanks to the large number of functional gadgets available through the use of
the ADG, we were able to make the attack stealthy, by restoring the nominal
behavior of the application after the attack is completed~--- a property that
we did not found in previous code-reuse attacks.

New challenges to increase practicability of JOP attacks include assistance in 
gagdet finding and even automated chain building.
There is a very impressive body of research on ROP chain
building~\cite{Vishnyakov21}, that would be a good basis to build up automated
testing frameworks for RISC-V application vulnerabilities to JOP.
Likewise, studies like the one presented in this article will enable the
development of better and more efficient countermeasures for the RISC-V
architecture against JOP attacks and enhance {control-flow integrity}
in general.

\subsubsection{Acknowledgements} This work is partly supported by the French
research agency (ANR) under the grant ANR-21-CE-39-0017.
We thank Franck Viguier for his contribution to a preliminary version 
of this work.

\bibliographystyle{splncs04}
\bibliography{bibliography}

\begin{thebibliography}{10}
\providecommand{\url}[1]{\texttt{#1}}
\providecommand{\urlprefix}{URL }
\providecommand{\doi}[1]{https://doi.org/#1}

\bibitem{abadi05}
Abadi, M., Budiu, M., Erlingsson, U., Ligatti, J.: Control-flow integrity. In:
  the 12th ACM Conference on Computer and Communications Security. p.
  340–353. CCS '05, ACM (2005). \doi{10.1145/1102120.1102165}

\bibitem{joprocket21}
Brizendine, B., Babcock, A.: Pre-built {JOP} chains with the {JOP} {ROCKET}:
  Bypassing {DEP} without {ROP}. Black Hat Asia (2021)

\bibitem{Burow17}
Burow, N., Carr, S.A., Nash, J., Larsen, P., Franz, M., Stefan, B., Payer, M.:
  Control-flow integrity: Precision, security, and performance. ACM Comput.
  Surv.  \textbf{50}(1) (2017). \doi{10.1145/3054924}

\bibitem{Carlini14}
Carlini, N., Wagner, D.: {ROP} is still dangerous: Breaking modern defenses.
  In: the 23rd USENIX Conference on Security Symposium. p. 385–399. SEC'14,
  USENIX Association (2014)

\bibitem{Checkoway10}
Checkoway, S., Davi, L., Dmitrienko, A., Sadeghi, A.R., Shacham, H., Winandy,
  M.: Return-oriented programming without returns. pp. 559--572 (2010).
  \doi{10.1145/1866307.1866370}

\bibitem{Cheng11}
Chen, P., Xing, X., Mao, B., Xie, L., Shen, X., Yin, X.: Automatic construction
  of jump-oriented programming shellcode (on the x86). In: the 6th ACM
  Symposium on Information, Computer and Communications Security. p. 20–29.
  ASIACCS '11, ACM (2011). \doi{10.1145/1966913.1966918}

\bibitem{Cowan98}
Cowan, C., Pu, C., Maier, D., Hintony, H., Walpole, J., Bakke, P., Beattie, S.,
  Grier, A., Wagle, P., Zhang, Q.: Stackguard: Automatic adaptive detection and
  prevention of buffer-overflow attacks pp.~5--5 (1998)

\bibitem{heapsafe22}
De, A., Ghosh, S.: Heapsafe: Securing unprotected heaps in {RISC-V}. In: 2022
  35th International Conference on VLSI Design and 2022 21st International
  Conference on Embedded Systems (VLSID). pp. 120--125 (2022).
  \doi{10.1109/VLSID2022.2022.00034}

\bibitem{Erdodi13}
Erdődi, L.: Finding dispatcher gadgets for jump oriented programming code
  reuse attacks. In: 2013 IEEE 8th International Symposium on Applied
  Computational Intelligence and Informatics (SACI). pp. 321--325 (2013).
  \doi{10.1109/SACI.2013.6608990}

\bibitem{gilles2022}
Gilles, O., Viguier, F., Kosmatov, N., {Gracia P\'erez}, D.: {Control-Flow
  Integrity at RISC: Attacking {RISC-V} by Jump-Oriented Programming} (2022),
  \url{https://arxiv.org/abs/2211.16212}

\bibitem{gu2020riscvrop}
Gu, G., Shacham, H.: No {RISC} no reward: Return-oriented programming in
  {RISC-V} (2020), \url{https://arxiv.org/pdf/2007.14995.pdf}

\bibitem{morpheus21}
Harris, A., Verma, T., Wei, S., Biernacki, L., Kisil, A., Aga, M.T., Bertacco,
  V., Kasikci, B., Tiwari, M., Austin, T.: Morpheus {II}: A {RISC-V} security
  extension for protecting vulnerable software and hardware. In: 2021 IEEE
  International Symposium on Hardware Oriented Security and Trust (HOST). pp.
  226--238 (2021). \doi{10.1109/HOST49136.2021.9702275}

\bibitem{Jaloyan20}
Jaloyan, G.A., Markantonakis, K., Akram, R.N., Robin, D., Mayes, K., Naccache,
  D.: Return-oriented programming on {RISC-V}. In: the 15th ACM Asia Conference
  on Computer and Communications Security. p. 471–480. ASIA CCS '20, ACM
  (2020). \doi{10.1145/3320269.3384738}

\bibitem{Nurmukhametov21}
Nurmukhametov, A., Vishnyakov, A., Logunova, V., Kurmangaleev, S.: {MAJORCA}:
  Multi-architecture {JOP} and {ROP} chain assembler. In: 2021 Ivannikov Ispras
  Open Conference (ISPRAS). pp. 37--46 (2021).
  \doi{10.1109/ISPRAS53967.2021.00011}

\bibitem{Palmiero18}
Palmiero, C., Di~Guglielmo, G., Lavagno, L., Carloni, L.P.: Design and
  implementation of a dynamic information flow tracking architecture to secure
  a {RISC-V} core for iot applications. In: 2018 IEEE High Performance extreme
  Computing Conference (HPEC). pp.~1--7 (2018). \doi{10.1109/HPEC.2018.8547578}

\bibitem{Sadeghi14}
Sadeghi, A.A., Aminmansour, F., Shahriari, H.R.: Tazhi: A novel technique for
  hunting trampoline gadgets of jump oriented programming (a class of code
  reuse attacks). In: 2014 11th International ISC Conference on Information
  Security and Cryptology. pp. 21--26 (2014). \doi{10.1109/ISCISC.2014.6994016}

\bibitem{Shacham07}
Shacham, H.: The geometry of innocent flesh on the bone: Return-into-libc
  without function calls (on the {X86}). In: the 14th ACM Conference on
  Computer and Communications Security. p. 552–561. CCS '07, ACM (2007).
  \doi{10.1145/1315245.1315313}

\bibitem{Shacham05}
Shacham, H., Page, M., Pfaff, B., Goh, E., Modadugu, N., Boneh, D.: On the
  effectiveness of address-space randomization. In: the 11th ACM Conference on
  Computer and Communications Security. p. 298–307. CCS '04, ACM (2004).
  \doi{10.1145/1030083.1030124}

\bibitem{Solardesigner97}
{Solar Designer}: Getting around non-executable stack (and fix). (1997),
  \url{https://seclists.org/bugtraq/1997/Aug/63}

\bibitem{Vishnyakov21}
Vishnyakov, A., Nurmukhametov, A.: Survey of methods for automated code-reuse
  exploit generation. Programming and Computer Software  \textbf{47},  271--297
  (2021). \doi{10.1134/S0361768821040071}

\bibitem{Younan04codeinjection}
Younan, Y., Joosen, W., Piessens, F.: Code injection in {C} and {C++}: A survey
  of vulnerabilities and countermeasures. Tech. rep., Departement
  Computerwetenschappen, Katholieke Universiteit Leuven (2004),
  \url{https://www.cs.kuleuven.be/publicaties/rapporten/cw/CW386.pdf}

\bibitem{zaruba2019cost}
{Zaruba}, F., {Benini}, L.: The cost of application-class processing: Energy
  and performance analysis of a {Linux}-ready {1.7-GHz} {64-Bit} {RISC-V} core
  in 22-nm {FDSOI} technology. IEEE Transactions on Very Large Scale
  Integration (VLSI) Systems  \textbf{27}(11),  2629--2640 (2019).
  \doi{10.1109/TVLSI.2019.2926114}

\end{thebibliography}

\end{document}